\documentclass[useAMS,usenatbib]{mn2e}

\usepackage{natbib, aas_macros}
\usepackage[dvips]{graphicx}
\usepackage{wrapfig}
\graphicspath{{./figs/}}
\usepackage{color}
\usepackage{amsmath}
\usepackage{amssymb}

\newcommand{\Msun}{M_{\odot}}

\newcommand{\Zsun}{Z_{\odot}}
\newcommand{\fesc}{f_{\rm esc}}

\newcommand{\Lsun}{L_{\odot}}

\newcommand{\MBH}{\rm M_{\rm{BH}}}

\newcommand{\Mstar}{\rm M_{\rm{star}}}

\newcommand{\lya}{\rm {Ly{\alpha}}}

\newcommand{\Msunyr}{\rm M_{\odot}~ yr^{-1}}

\newcommand{\Tb}{\delta T_{\rm b}}
\newcommand{\Ts}{T_{\rm S}}
\newcommand{\Tcmb}{T_{\rm CMB}}

\newcommand{\Tgas}{T_{\rm gas}}
\newcommand{\tevo}{t_{\rm evo}}

\title[Distinctive 21 cm Structures of the First Stars, Galaxies, and Quasars]
{Distinctive 21 cm Structures of the First Stars, Galaxies, and Quasars}
\author[Yajima \& Li]
{Hidenobu Yajima$^{1}$\thanks{E-mail: yajima@roe.ac.uk (HY)} and Yuexing Li$^{2, 3}$ 
\\
$^{1}$ SUPA\thanks{Scottish Universities Physics Alliance}, Institute for Astronomy, University of Edinburgh, Royal Observatory, Edinburgh, EH9 3HJ, UK\\
$^{2}$Department of Astronomy and Astrophysics, Pennsylvania State University,
525 Davey Lab, University Park, PA 16802, USA\\
$^{3}$Institute for Gravitation and the Cosmos, The Pennsylvania State University, University Park, PA 16802, USA
}
\begin{document}

\date{Accepted ?; Received ??; in original form ???}

\pagerange{\pageref{firstpage}--\pageref{lastpage}} \pubyear{2008}

\maketitle

\label{firstpage}

%----------------------------------------------------------------------
%
% Abstract
%
%----------------------------------------------------------------------
\begin{abstract}

Observations of the redshifted 21 cm line with upcoming radio telescopes promise to transform our understanding of the cosmic reionization. To unravel the underlying physical process, we investigate the 21 cm structures of three different ionizing sources, Population (Pop) III stars, the first galaxies, and the first quasars, by using radiative transfer simulations that include both ionization of neutral hydrogen and resonant scattering of $\lya$ photons. We find that Pop III stars and quasars  produce a smooth transition from an ionized and hot state to a neutral and cold one, owing to their hard spectral energy distribution with abundant ionizing photons, in contrast to the sharp transition in galaxies. Furthermore, $\lya$ scattering plays a dominant role in producing the 21 cm signal as it determines the relation between hydrogen spin temperature and gas kinetic temperature. This effect, also called Wouthuysen-Field coupling, depends  strongly on the ionizing source. It is the strongest around galaxies, where the spin temperature is highly coupled to that of  the gas, resulting in extended absorption troughs in the 21 cm brightness temperature. 
On the other hand, in the case of Pop III stars, the 21 cm signal shows both emission and absorption regions around a small H{\sc ii} bubble.
For quasars, a large emission region in the 21 cm signal is produced, and the absorption region decreases as the size of the H{\sc ii} bubble becomes large due to the limited traveling time of photons. 
We predict that future surveys from large radio arrays such as MWA, LOFAR and SKA may be able to detect the 21 cm signals of primordial galaxies and quasars, but not likely Pop III stars due to its small angular diameter. 

\end{abstract}

%----------------------------------------------------------------------
%
% Keywords
%
%----------------------------------------------------------------------
\begin{keywords}
radiative transfer -- diffuse radiation -- galaxies: evolution -- galaxies: formation -- galaxies: high-redshift -- quasars: supermassive black holes
\end{keywords}

%----------------------------------------------------------------------
%
% Section 1: Introduction
%
%----------------------------------------------------------------------
\section{Introduction}

The epoch of reionization (EoR), during which high energy photons produced by the first luminous objects reionized the neutral hydrogen in the intergalactic medium (IGM), was an important milestone in cosmic history \citep{Loeb01}. The latest measurements from the Planck satellite suggest that  the Universe was  reionized at redshift $z \sim 11$ \citep{Planck13}, in agreement with the seven-year results of  Wilkinson Microwave Anisotropy Probe (WMAP) \citep{Komatsu11}, while studies of Gunn-Peterson absorption \citep{Gunn65} of high-redshift quasars (QSOs) suggest that reionization began as early as $z \sim 15$ and ended at $z \sim 6$ \citep{Fan06a}. The reionization history strongly constrains not only the formation of the first generation of galaxies and QSOs, but also their feedback and impacts on structure formation at later times \citep{Bromm11}. 

 Over the past few years, impressive progress has been made in detecting distant objects. Recent observations using both broad-band colors (e.g., \citealt{Bouwens12, Ellis13}) and narrow-band Ly$\alpha$ emission (e.g., \citealt{Ouchi10, Kashikawa11}) have detected hundreds of galaxies at $z \gtrsim 6$.  Meanwhile, over two dozen luminous QSOs have been detected at $z \sim 6$ (e.g., \citealt{Fan06b, Willott10a, Mortlock11}). While it is generally believed that early star-forming galaxies played an important role in reionizing the Universe  \citep{Robertson10}, the ionization process and the actual contributions from different ionizing sources remain poorly understood. 
 
The 21 cm hyperfine line of neutral hydrogen has been proposed as a powerful tool to probe the EoR, as it traces the thermal history of the IGM and the ionization structures (e.g., \citealt{Morales10, Pritchard12}). Recent advances in radio instrumentation and techniques will soon make it possible to measure the highly redshifted 21 cm line from gas during the first billion years after the Big Bang, as  a number of radio interferometers are currently being built or planned, such as Murchison Widefield Array  \citep[MWA; ][]{Lonsdale09}, the LOw Frequency ARray \citep[LOFAR; ][]{Harker10}, the Precision Array to Probe the Epoch of Reionization  \citep[PAPER; ][]{Parsons10}, %the 21 cm Array (21CMA),
 the Giant Meter-wave Radio Telescope \citep[GMRT; ][]{Paciga11}, and the Square Kilometre Array \citep[SKA; ][]{Dewdney09}. 

In order to understand and interpret observations from these instruments, it is critical to understand the 21 cm structures from different ionizing sources. It has been suggested by advanced numerical simulations that the first stars (so-called ``Population III'' stars) started to form in mini-halos of $10^{5 - 6}\, \Msun$ as early as $z \sim 30$ \citep{Abel02a, Bromm02, Yoshida08, Turk09, Clark11}. The first galaxies are believed to have formed in low-metallicity halos via gas accretion or mergers around $z \sim 15$ thanks to feedback and metal enrichment from Population (Pop) III stars \citep{Wise07, Wise08a, Wise08b, Greif10, Pawlik11, Wise12a}. Meanwhile, the massive black holes (BHs) may have formed from remnants of massive Pop III stars or direct collapse of gas clumps or supermassive stars (e.g., \citealt{Volonteri10}, leading to the emergence of the first QSOs \citep{Li07, DiMatteo08, Sijacki09, DiMatteo12}. These objects provided strong  UV radiation that reionized the neutral hydrogen. 

To date, a number of theoretical works have studied the 21 cm signals of Pop III stars \citep{Chen04, Chen08, Tokutani09}, galaxies \citep{Furlanetto06, McQuinn06, Kuhlen06, Mellema06b, Semelin07, Baek10, Vonlanthen11, Mesinger11, Iliev12}, and QSOs \citep{Wyithe05, Wyithe07, Geil08, Alvarez10, Datta12, Majumdar12}. The 21 cm signal can be emission or absorption against the cosmic microwave background (CMB), if spin temperature ($\Ts$) is above or below the CMB temperature ($\Tcmb$). The scattering process of $\lya$ photons is the main mechanism which causes $\Ts$ decouple from $\Tcmb$ via the Wouthuysen-Field effect \citep{Wouthuysen52, Field58, Hirata06}. Therefore, it is crucial to take into account $\lya$ scattering in the calculation of 21 cm signals. The $\lya$ radiation field in  a moving medium like the Hubble flow is difficult to estimate analytically, because the diffuse approximation is no longer valid in a partially ionized or moving medium with higher relative velocity \citep{Loeb99}, and so must be determined through detailed radiative transfer (RT) calculations. However, despite the importance of the Wouthuysen-Field effect, the $\lya$ radiation field was simply modeled or completely ignored in most of the previous work; only a few actually included $\lya$ RT in the estimate of $\Ts$ \citep[e.g.,][]{Baek10, Vonlanthen11}. In particular, \cite{Baek09} and \cite{Baek10} combined cosmological simulations and post-processing $\lya$ RT to calculate the 21 cm signals from UV and X-ray sources. They found that the absorption phase of the 21 cm survives throughout the EoR  even in the presence of strong X-ray sources, and that the brightness temperature fluctuation of the 21 cm signal evolves strongly with redshift, with a higher amplitude in the early reionization phase. 

In order to investigate the Wouthuysen-Field effect from different ionizing sources and the resulting 21 cm structures, we carry out here a comparative study of the ionization history of three types of sources: Pop III stars, primordial galaxies and QSOs. We perform RT calculations, which include both ionization and $\lya$ scattering, on individual sources embedded in the IGM. Different intrinsic spectral energy distributions (SEDs) appropriate for each source type are used. We follow the evolution of the ionization structures and temperature to derive the 21 cm signals.  

The paper is organized as follows. In \S2, we describe the model and methodology underlying our simulations. In \S3, we present the results, which include the structure of ionization, temperatures and the 21 cm signal around different sources, and their time evolution. We discuss in \S4 the detectability of the 21 cm structures of these sources by upcoming facilities such as MWA, LOFAR and SKA, and we summarize in \S5. 

%----------------------------------------------------------------------
%
% Section 2:  Model and Method
%
%----------------------------------------------------------------------
\section{Model \& Method}

In this work, we carry out RT calculations that include ionization and $\lya$ scattering on three types of ionizing sources embedded in the IGM: Pop III stars, galaxies, and QSOs. The IGM is modeled as spherical shells around the source. The boundary of the sphere is set to $R=7 R_{\rm S}$, where 
$R_{\rm S}$ is the radius of Str\"{o}mgren sphere, and it is linearly divided into 300 bins. 

For each source type, we consider a range of masses in the redshift range $z=7-20$. For Pop III stars, 
$\Mstar = 2\times10^{2} - 2\times10^{5}~\rm \Msun$; for galaxies, $\Mstar = 2\times10^{6} - 2\times10^{10}~\rm \Msun$,  and for QSOs, $\Mstar = 2\times10^{8} - 2\times10^{12}~\rm \Msun$, 
where  $\Mstar$ is total stellar mass. For Pop III stars with $\Mstar > 10^{2}~\rm \Msun$,  we assume multiple star formation of $100~\rm \Msun$ in a halo \citep[e.g.,][]{Greif12}, or a cluster of halos with Pop III stars. 

\subsection{The 21 cm Signal}

The fluctuation in 21 cm intensity (or fluctuation of brightness temperature) from different regions of the IGM at a given redshift $z$ depends sensitively on the gas properties, including density, velocity gradients, gas temperature, gas spin temperature and ionization state \citep{Furlanetto06}. We follow the procedure of previous works \citep[e.g.,][]{Furlanetto06, Baek09} which calculate the fluctuation of the brightness temperature as 

\begin{eqnarray}
\delta T_{\rm b} = 28.1~ \chi_{\rm HI} (1+\delta) \left( \frac{1+z}{10} \right)^{1/2} \frac{\Ts - T_{\rm CMB}}{\Ts}~~  {\rm mK},
% &\times& \left( \frac{\Omega_{\rm b}}{0.042}\frac{h}{0.73} \right) \left( \frac{0.24}{\Omega_{\rm m}} \right)^{1/2} \left(  \frac{1-Y_{\rm p}}{1-0.248} \right),
\end{eqnarray}
where $\chi_{\rm HI}$ is the neutral hydrogen fraction, $\delta$ is over density, $\Ts$ and $\Tcmb$ are the gas spin and CMB temperature respectively. The IGM is assumed to be uniform, i.e., $\delta=0$, as recent simulations suggest small clumpiness in the IGM at high redshift \citep{Pawlik09b}. The contribution of the gradient of the proper velocity is not considered in this work.

The gas spin temperature is controlled by Thomson scattering of CMB photons, $\lya$ photon pumping, and collisions by the gas particles, as formulated by \cite{Furlanetto06}: 
\begin{equation}
\label{eq:ts}
\Ts^{-1} = \frac{\Tcmb^{-1} + x_{\rm C}\Tgas^{-1} + x_{\rm \alpha}T_{\rm C}^{-1}}{1 + x_{\rm C} + x_{\rm \alpha}},
\end{equation}
where $T_{\rm C}$ is the color temperature of the $\lya$ line, $\Tgas$ is the kinetic temperature of the gas, $x_{\rm \alpha}$ and $x_{\rm C}$ are the coupling coefficients by $\lya$ photon scattering and gas collision respectively, which are calculated for each spherical shell from RT simulations of ionizing and $\lya$ photons as follows
\begin{eqnarray}
x_{\rm \alpha} & = & \frac{4 P_{\rm \alpha} T_{\star}}{27 A_{10} \Tcmb}, \\
x_{\rm C} &  = & \frac{T_{\star}}{A_{10} \Tcmb}(C_{\rm H} + C_{\rm p} + C_{\rm e}). 
\end{eqnarray}
where  $T_{\star} = 0.068~\rm K$, $A_{10} = 2.85\times 10^{-15}~\rm s^{-1}$ is the spontaneous emission factor of the 21 cm transition, $P_{\alpha}$ is the number of scatterings of $\lya$ photons per atom per second, and $C_{\rm H}$,  $C_{\rm p}$ and $C_{\rm e}$ are the de-excitation rates due to collision with neutral atoms, protons, and electrons, respectively. We use the fitting formula given by \citet{Liszt01} and \citet{Kuhlen06} for the de-excitation rates. 

Since $T_{\rm C}$ quickly settles into $\Tgas$ owing to the recoil effect of $\lya$ photon scattering, $T_{\rm C} = \Tgas$ is assumed in our calculations. The spin temperature depends sensitively on the coupling due to $\lya$ scattering and collision. When the coupling is strong, it decouples from the CMB temperature and becomes $\Ts \sim \Tgas$.

\subsection{Intrinsic SEDs of Ionizing Sources}

We consider different SEDs for the three types of sources, PopIII stars, galaxies, and QSOs in our calculations. For PopIII stars, we assume a black body spectrum, and follow the formulae from \cite{Bromm01} to calculate the effective temperature and bolometric luminosity of a star with mass $M$:  
\begin{eqnarray}
T_{\rm PopIII}^{\rm eff} &=& 1.1 \times 10^{5} (\frac{M}{100~\Msun})^{0.025} ~{\rm K}, \\
 L_{\rm PopIII}^{\rm bol} &=& 10^{4.5} \frac{M}{\Msun} \Lsun.
\end{eqnarray}
Haloes hosting Pop III stars can be metal-enriched quickly due to type-II or pair-instability supernovae \citep{Wise12a, Johnson13}.
As a result, metallicity can exceed the threshold of Pop III star formation, $Z/Z_{\rm \odot} \sim 10^{-3.5}$ \citep{Bromm01b},
and then Pop II stars form in halos \citep[e.g.,][]{Wise12a}.
On the other hand, recent simulations showed that multiple Pop III stars form in each halo \citep{Clark11, Greif12}, 
and haloes are clustered close together \citep{Umemura12}. 
In addition, if the mass of Pop III stars is $\gtrsim 260~\Msun$, they collapse directly into black holes \citep{Heger02}, 
as a result, the halo can keep making Pop III stars even after several Myr. 
Hence, in this work, we assume Pop III halos which consist of Pop III stars alone. 
Effects of mixed systems of Pop III and Pop II stars on 21 cm signal will be investigated by detailed hydrodynamics simulations in future work.

The primordial galaxies in our model are assumed to consist of only young, metal-poor stars which follow a Salpeter initial mass function (IMF) \citep{Salpeter55}. We assume a stellar age of 10 Myr and a metallicity of $Z = 0.1~\Zsun$, which are consistent with observations of high-redshift $\lya$ emitting galaxies (e.g., \citealt{Lai07, Finkelstein09b, Ota10, Nakajima12}). The SEDs of galaxies are then generated using the stellar population synthesis code {\sc P\'{E}GASE} v2.0 \citep{Fioc97} for a given stellar mass with the above parameters of stellar age, metallicity and IMF.  
In practice, the SED changes with evolution time. 
However, the star formation history of high-redshift galaxies is as yet poorly understood. 
In this work, for simplicity, we use the above SED with constant age, metallicity and stellar mass. 
Note that, however, even if we consider an evolving SED with a constant star formation model like ${\rm SFR} = (M_{\rm star} / 10^{8}~\Msun) ~ \Msunyr$ \citep[e.g.,][]{Ono10}, 
the differences of sizes of H{\sc ii} bubbles are within a factor of $\sim 2$. 

The SED of QSOs is estimated by adding a broken power-law spectrum of massive black holes  to the galaxy's SED. 
The power-law spectrum is assumed, following \citet{Laor93, Marconi04, Hopkins07}, to be
\begin{equation}
\nu L_{\nu} \propto \begin{cases}
\nu^{1.2} & {\rm for} ~{\rm log}\; (\nu/{\rm Hz}) < 15.2\\
\nu^{-1.2}
& {\rm for} ~{\rm log}\; (\nu/{\rm Hz}) > 15.2. 
\end{cases}
\end{equation}
Recent observations of the most distant QSOs show that the BH accretion rates are close to the Eddington limit \citep{Willott10b},  and simulations of early QSOs by \cite{Li07} showed that the mass ratio of BHs to stars in the hosts is $\sim 10^{-3}$, similar to that of local galaxies (e.g., \citealt{Marconi03, Rix04}). Therefore, in our QSO model, the BH mass is set to be $\MBH \sim 10^{-3} \times \Mstar$, and its luminosity is derived assuming an Eddington limit.
For ionization and heating, we integrate the above SED up to $40~\rm KeV$. Contribution of higher-energy photons $E > 40~\rm KeV$
is negligible because of the small energy fraction and low ionization cross section.
In addition, free electrons made by X-ray ionization are highly energetic, 
and they make $\lya$ photons via collisions with neutral hydrogen. 
Roughly 40 per cent of X-ray energy can be used for the production of $\lya$ photons, and they can enhance the 21 cm signal \citep{Chen08}.
We also include the additional $\lya$ photons by converting 40 per cent of X-ray energy at $E > 0.1~\rm KeV$. 

\begin{figure}
\begin{center}
\includegraphics[scale=0.4]{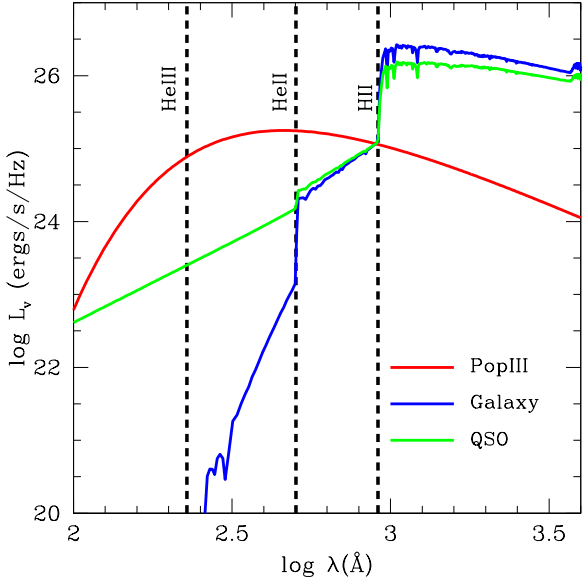}
\caption{
The intrinsic SEDs of a Pop III star (red curve), a galaxy  (blue curve), and a QSO (green curve).  The SEDs of the Pop III star and QSO are normalized by the flux of galaxy with $\Mstar = 10^{6}~\rm \Msun$ at $900~\rm \AA$. The three vertical dashed lines indicate the wavelengths of He{\sc iii}, He{\sc ii} and H{\sc ii}, respectively. 
}
\label{fig:sed}
\end{center}
\end{figure}

The SEDs of these three types of sources are shown in Figure~\ref{fig:sed}. Pop IIII stars have a hard SED with a peak at $\lambda \sim 300~\rm \AA$.  
The SED of galaxies drops sharply at wavelengths shortward of the Lyman limit $\lambda \sim 912~\rm \AA$, while that of QSO has a power-law tail from X-rays to the Lyman limit owing to radiation from an accreting BH. As we will show later, the difference in the SEDs at $\lambda \lesssim 912~\rm \AA$  would result in  different  ionization structures of hydrogen by the three ionizing sources. On the other hand, both galaxies and QSOs have much higher continuum flux between $\lya$ and Ly$\beta$ frequencies %$\lya$ flux at $\lambda = 1216~\rm \AA$ 
than Pop III stars, which would lead to stronger effects of $\lya$ scattering.

\subsection{Ionization and Heating of the IGM}

We calculate the ionization of hydrogen and helium in the IGM by central sources using one-dimensional RT of ionizing photons. 
Recent simulations showed that the escape fraction of ionizing photons ($\fesc$) of high-redshift galaxies or Pop III halos can be $\gtrsim 0.5$
 \citep{Yoshida07, Wise09, Yajima11, Paardekooper13}.
Hence, we simply assume that half of ionizing photons are absorbed by the interstellar medium (ISM) and the rest escape from halos and ionize IGM.
About 0.68 of ionizing photons absorbed by ISM or IGM are converted to $\lya$ photons via recombination processes.
Here we ignore detailed ionization structure in halos of which the spatial scale is much smaller than the typical size of H{\sc ii} bubbles in the IGM. In addition,
%In this work, 
we focus on the early phase when the radiation field is localized, hence the radiation background from external sources is not included.
The time evolution of the ionization of hydrogen and helium is estimated by the following equations:
\begin{eqnarray}
\frac{d\chi_{\rm HI}}{dt} & = & - \Gamma^{\gamma}_{\rm HI} -  \Gamma^{\rm C}_{\rm HI}\chi_{\rm HI}n_{\rm e} + \alpha_{\rm B}^{\rm H} \chi_{\rm HII} n_{\rm e}, \\
\frac{d\chi_{\rm HeI}}{dt} & =  & - \Gamma^{\gamma}_{\rm HeI} - \Gamma^{\rm C}_{\rm HeI}\chi_{\rm HeI}n_{\rm e} + \alpha_{\rm B}^{\rm HeI} \chi_{\rm HeII} n_{\rm e}, \\
\frac{d\chi_{\rm HeII}}{dt} & =  & \Gamma^{\gamma}_{\rm HeI} - \Gamma^{\gamma}_{\rm HeII} 
				        +  \Gamma^{\rm C}_{\rm HeI}\chi_{\rm HeI}n_{\rm e}  - \Gamma^{\rm C}_{\rm HeII}\chi_{\rm HeII}n_{\rm e} \cr  \nonumber \\
				        &  & - \alpha_{\rm B}^{\rm HeI} \chi_{\rm HeII} n_{\rm e} + \alpha_{\rm B}^{\rm HeII}\chi_{\rm HeIII} n_{\rm e}. 
\end{eqnarray}
Here, $\chi_{\rm HI}$,  $\chi_{\rm HeI}$, $\chi_{\rm HII}$,  and $\chi_{\rm HeII}$ are the ionization fractions of neutral hydrogen, neutral helium, ionized hydrogen and ionized helium, respectively, $n_{\rm e}$ is the electron density, $\alpha_{\rm B}$ is the case B recombination rate, $\Gamma^{\rm C}$ is the collisional ionization rate. We use the fitting formula of \citet{Gnedin97} for $\alpha_{\rm B}$, and that of \citet{Cen92} for $\Gamma^{\rm C}$. The photo-ionization rate $\Gamma^{\gamma}$ in each shell is estimated by
\begin{eqnarray}
\Gamma^{\gamma}_{\rm HI} = \frac{1}{n_{\rm H} V_{\rm shell}} 
 \int_{\nu_{\rm limit}}^{\infty} \frac{L(\nu)}{h\nu} e^{-\tau(\nu)}\left(1-e^{-\Delta \tau_{\rm HI}(\nu)}\right) d\nu
\end{eqnarray}
where $V_{\rm shell}$ is the volume of the gas shell, $\nu_{\rm limit}$ is the Lyman limit frequency,
$\tau$ is the optical depth from the central source to the shell, and $\Delta \tau$ is optical depth of the shell.
The optical depth is calculated by
\begin{eqnarray}
\tau(\nu) &=& n_{\rm H} \sigma_{\rm HI}(\nu) \int_{0}^{r} dr^{'} \chi_{\rm HI}(r^{'}) 
 + n_{\rm He}\sigma_{\rm HeI}(\nu) \int_{0}^{r} dr^{'} \chi_{\rm HeI}(r^{'})  \cr
 & & + ~ n_{\rm He}\sigma_{\rm HeII}(\nu) \int_{0}^{r} dr^{'} \chi_{\rm HeII}(r^{'})
\end{eqnarray}
where $\sigma$ is the ionization cross section. We use the fitting formula from \citet{Cen92}, $\sigma_{\rm HI} = 6.3\times10^{-18} (\nu / \nu_{\rm limit}^{\rm HI})^{-3}$,
$\sigma_{\rm HeI} = 7.2 \times 10^{-18} [1.66 (\nu / \nu_{\rm limit}^{\rm HeI})^{-2.05} + 0.66 (\nu / \nu_{\rm limit}^{\rm HeI})^{-3.05}]$,
$\sigma_{\rm HeII} = 1.58 \times 10^{-18} (\nu / \nu_{\rm limit}^{\rm HeII})^{-3}$. 

The evolution of ionization would lead to change of gas temperature in each shell with time as follows,
\begin{equation}
\frac{d\Tgas}{dt} = \frac{2}{3k_{\rm B}n} \left[ k_{\rm B}\Tgas\frac{dn}{dt} + H - \Lambda \right]
\end{equation}
where $k_{\rm B}$ is the Boltzmann constant, $n$ is the total number density of hydrogen and helium, $H$ is the heating rate, and $\Lambda$ is the cooling rate.
We follow the procedures of \cite{Maselli03} to calculate the heating and cooling rates: we consider photo-heating for $H$, while for $\Lambda$ , we include recombination, collisional ionization and excitation cooling processes. The upper limit of the gas temperature is set to $10^{5}~\rm K$, and heating by $\lya$ photon scattering is not included, as it was shown to be insignificant by \citet{Chen04}.  

We follow the evolution of ionization and temperature up to $10^{8}~\rm yr$, and we calculate the $\lya$ radiation field and the 21 cm signal of the snapshots at $t_{\rm evo} = 10^{6}, 10^{7}$ and $10^{8}$ yr. Our fiducial runs use the snapshot at $\tevo = 10^{7}$ yr at redshift $z=10$.

\subsection{Radiative Transfer of $\lya$ Photons}
\label{sec:radiation}

%%%%%%%%%%%%

The RT of $\lya$ photons in IGM is a highly complex process. It depends strongly on the $\lya$ resonant scattering, and on the density distribution and ionization state of the medium. The frequency change resulting from the scattering is difficult to estimate analytically. \cite{Yajima12b} has developed a three-dimensional, Monte Carlo $\lya$ RT code which couples the continuum and ionization of hydrogen. Here we use the 1-D version of the code to numerically simulate the $\lya$ RT in spherical shells   of the IGM.  

In the scattering process, the outgoing frequency in the laboratory frame $\nu^{\rm out}$ is calculated as
\begin{equation}
%x_{f} = x_{i} - u_{\parallel} + \mathbf{n}_{f}\cdot\mathbf{u} 
\frac{\nu^{\rm out} - \nu_{0}}{\Delta \nu_{\rm D}} = \frac{\nu^{\rm in} - \nu_{0}}{\Delta \nu_{\rm D}}
- \frac{v_{\rm a} \cdot k_{\rm in}}{v_{\rm th}} + \frac{v_{\rm a} \cdot k_{\rm out}}{v_{\rm th}}
\end{equation}
where $\nu^{\rm in}$ is the incoming frequency in the rest frame of scattering medium, $\nu_{0}$ is the line center frequency $\nu_{0} = 2.466 \times 10^{15}$ Hz,
$v_{\rm a}$ is the atom velocity, $\Delta \nu_{\rm D} = (v_{\rm th} / c) \nu_{0}$ corresponds to the Doppler frequency width, 
$v_{\rm th}$ is the velocity dispersion of the Maxwellian distribution describing the thermal motions, i.e., $v_{\rm th} = (2 k_{\rm B} T / m_{\rm H})^{1/2}$, and $k_{\rm in}$ and $k_{\rm out}$ are the incoming and outgoing propagation directions, respectively.

We consider $\lya$ photons from both continuum radiation of sources and recombination processes in ionized ISM and IGM.
The intrinsic $\lya$ luminosity is estimated by 
\begin{equation}
\label{eq:La}
\begin{split}
L_{\lya} = \int_{912~\; \rm \AA}^{1216~\; \rm \AA}  P_{\rm abs} f_{\rm conv}  L_{\lambda} d\lambda 
&+0.68 h \nu_{\alpha} (1-\fesc)\dot{N}_{\rm Ion}  \\
&+0.68 h \nu_{\alpha} \alpha_{\rm B}^{\rm H} n_{\rm HII} n_{\rm e} V_{\rm I}
\end{split}
\end{equation}
where the continuum spectrum is considered in the frequency range from the $\lya$ line ($1216~\rm \AA$) to Lyman limit ($912~\rm \AA$), 
$P_{\rm abs}$ is the absorption fraction in the calculation boxes, 
$f_{\rm conv}$ is the conversion fraction to $\lya$ photons via the cascade decay from higher quantum levels to the ground state \citep{Hirata06},
 and $\nu_{\alpha}$ is the $\lya$ frequency.
The $\lya$ from ISM is simply proportional to number of absorbed ionizing photons because of short recombination time-scale in ISM, while
that from IGM depends on the volume of the ionized bubble $V_{\rm I}$ and the density of the ionized hydrogen and electrons. 
Photons in this continuum range can be absorbed by neutral hydrogen which is then excited to higher levels of  the Lyman series, 
and some of hydrogen atoms can transit from the 2p to the 1s state by emitting $\lya$ photons.
Since we focus on the early phase of ionization up to $10^8$~yr in this work, which is shorter than the recombination time scale in IGM ($t_{\rm rec} \sim 4\times 10^{8}~\rm yr$), the ionized volume $V_{\rm I}$ in Equation~\ref{eq:La} is generally smaller than the Str\"{o}mgren sphere at ionization equilibrium. 
%Hence, the total $\lya$ emission is dominated by the continuum radiation from central sources.
Hence, the $\lya$ emission from IGM is much smaller than the other components. 
The $\lya$ emission from ISM is dominant for Pop III stars due to the high effective temperature, 
while the contribution of the continuum radiation is dominant for galaxies and QSOs. 

\begin{figure}
\begin{center}
\includegraphics[scale=0.4]{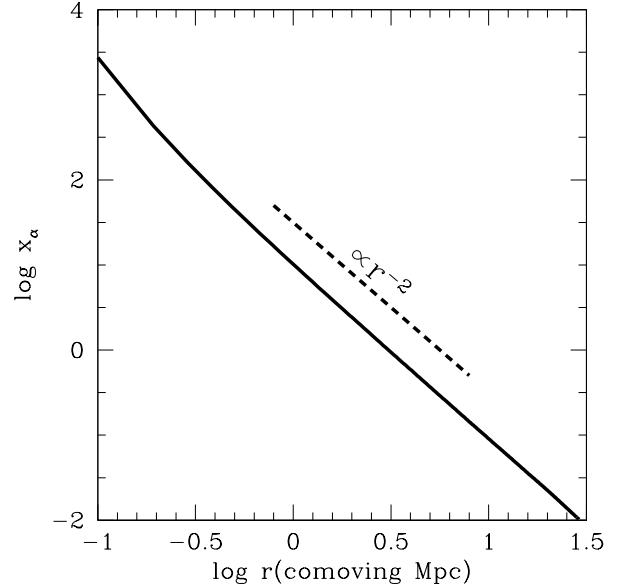}
\caption{
Profile of the coupling coefficient by continuum photons as a function of radial distance from a QSO of $\Mstar=2 \times 10^{8}~\Msun$ at $z=10$. 
Uniform IGM density on Hubble flow is assumed.
}
\label{fig:r2}
\end{center}
\end{figure}

We simulate the $\lya$ RT using the structure of  ionization and temperature at times $t_{\rm evo} = 10^{6}, 10^{7}$ and $10^{8}$~yr, by assuming that the travel time of the $\lya$ photons $t_{\rm travel} = t_{\rm evo}$. 
We divide the $\lya$ radiative transfer calculations into two parts. 
First, we calculate the RT of $\lya$ photons from the recombination process 
which are emitted from central sources.
These $\lya$ photons diffusely propagate outward. 
We performed a convergence test and found an optimal number of photon packets for the RT calculations, $N_{\rm p}=10^{5}$, of which the number of $\lya$ photons $ N_{\lya} = L_{\lya} / (h\nu_{\alpha} N_{\rm p})$. In these calculations, since we need to estimate the rate of scattering of $\lya$ photons in each shell precisely to derive the spin temperature, we can not use the ``core skipping'' technique in the $\lya$ RT which can accelerate the calculation, as used in most $\lya$ RT simulations \citep[e.g.,][]{Zheng02, Verhamme06, Dijkstra06, Laursen09a, Yajima12d, Yajima12e}. 
Hence, the calculations are very expensive even with $10^{5}$ photon packets. Each simulation took a few days running on 64 processors. 
Second, we consider continuum photons between $\lya$ and Lyman limit frequencies.
Total energy of these photons is larger than that of the recombination photons in the cases of galaxies and QSOs. 
Unlike the $\lya$ photons from the recombination process, the continuum photons can travel for long distances until they are absorbed at the Lyman series frequencies after cosmological redshifting \citep{Pritchard06, Vonlanthen11}.
For example, at $z \sim 10$, the photons for which the frequency is just below Ly$\beta$ can travel $\sim 40~\rm Mpc$ before the frequency is shifted to $\lya$. 
Once they are trapped, some fraction are converted to $\lya$ photons after the cascade decay \citep{Hirata06}
and experience numerous scatterings. 
\citet{Baek09} showed the number of scattering at the trapped place could be approximated by $N_{\rm sca} = 8 \times 10^{5} H(z=10)/H(z)$,
ignoring spatial diffusion.  
We use this approximation for the continuum photons. 
For uniform density and temperature, the scattering rate per atom by the continuum photons decreases with the $r^{-2}$ profile as the radial distance from sources becomes large \citep{Pritchard06, Chuzhoy07, Semelin07, Naoz08}.
As shown in Figure~\ref{fig:r2}, we also test the distribution of the coupling coefficient by continuum photons and show the roughly $r^{-2}$ distribution for a case of QSO of $\Mstar=2 \times 10^{8}~\Msun$ in uniform IGM on Hubble flow at $z=10$, which is similar to the test calculation of \citet{Semelin07} \citep[other test calculations were shown in][]{Yajima12b}.
The slight difference from the $r^{-2}$ profile is due to the not flat SEDs shown in Figure~\ref{fig:sed}.

%%%%%%%%%%%%%%%%%%%%%%%%%%%%%%%%%%%%%%%

%----------------------------------------------------------------------
%
% Section 3:  Results
%
%----------------------------------------------------------------------

\section{Results}

We performed a set of RT simulations on three types of ionizing sources, Pop III star, primordial galaxy and QSO, in the redshift range $z=7-20$. A wide range of masses were considered for these sources: $\Mstar = 2\times10^{2} - 2\times10^{5}~\rm \Msun$ for Pop III stars or cluster of Pop III stars,  $\Mstar = 2\times 10^{6} - 2\times10^{10}~\rm \Msun$ for galaxies,  and $\Mstar = 2\times10^{8} - 2\times10^{12}~\rm \Msun$ for QSOs. 

\subsection{Structures of Ionization and Temperature}
\label{sec:ionization}

\begin{figure}
\begin{center}
\includegraphics[scale=0.4]{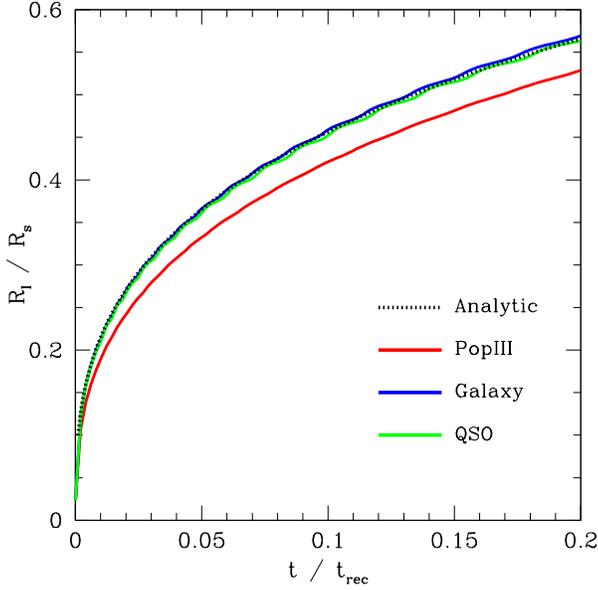}
\caption{
Propagation of the ionization front (normalized by radius of the Str\"{o}mgren sphere) as a function of time (normalized by recombination timescale), for a Pop III star ($\Mstar = 2 \times10^{2}~\rm \Msun$, red), a galaxy ($\Mstar = 2\times10^{6}~\rm \Msun$, blue), and a QSO ($\Mstar = 2\times10^{8}~\rm \Msun$, green), respectively. The black dotted line is the analytical solution from \citet{Spitzer78}.
}
\label{fig:ifront}
\end{center}
\end{figure}

\begin{figure}
\begin{center}
\includegraphics[scale=0.55]{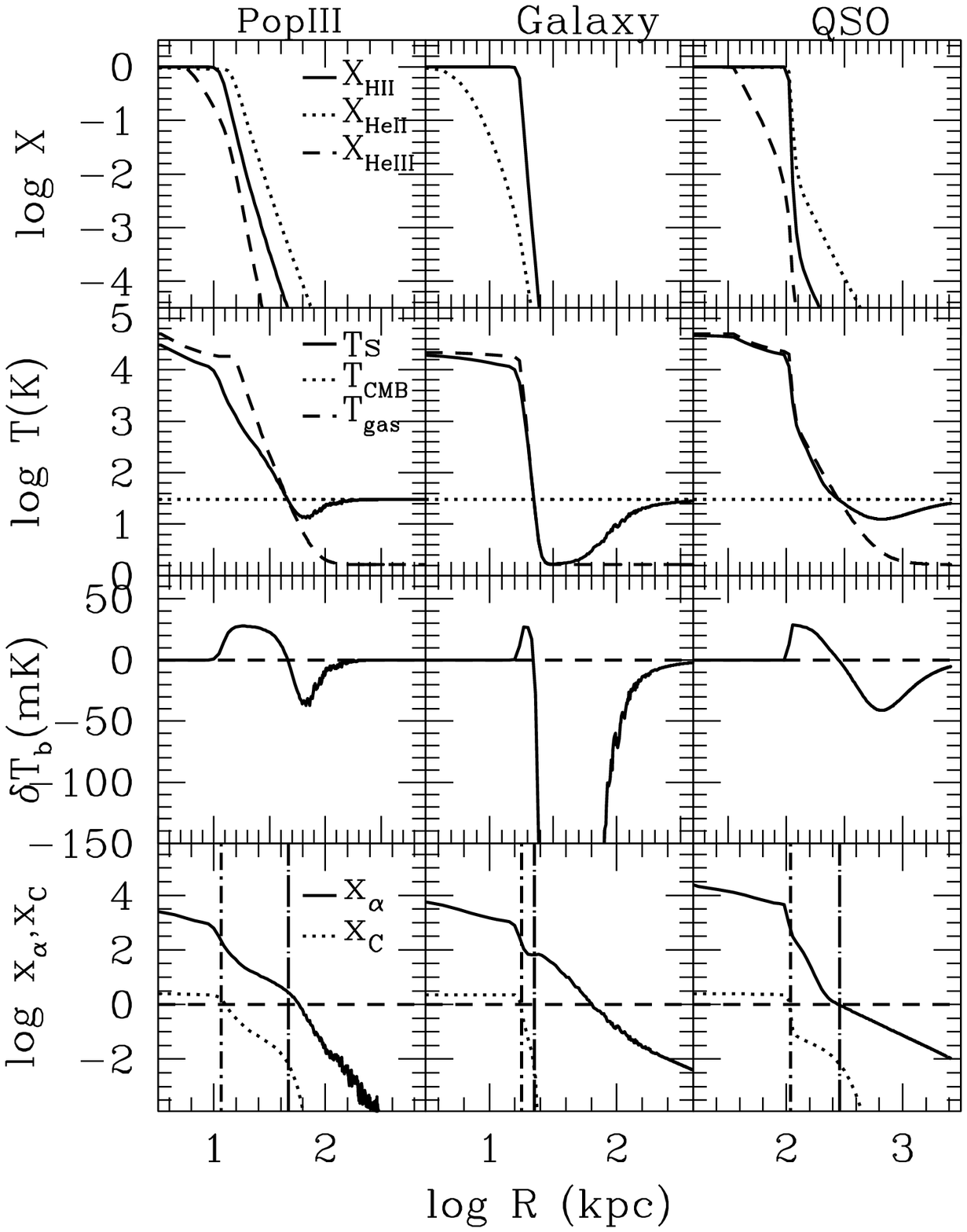}
\caption{
Structure of ionization, temperature, coupling coefficients by $\lya$ photons, and differential brightness temperature as a function of distance from the central source at redshift $z=10$ for a Pop III star ($\Mstar = 2\times10^{2}~\rm \Msun$, left column), a galaxy ($\Mstar = 2\times10^{6}~\rm \Msun$, middle column), and a QSO ($\Mstar = 2\times10^{8}~\rm \Msun$, right column), respectively. 
The top row panels show the fraction of H{\sc ii}, He{\sc ii}, and He{\sc iii} at $\tevo=10^{7}~\rm yr$, respectively.
The second row panels show the temperatures of spin, gas and CMB, respectively.
The third row panels are the fluctuations of  brightness temperature. 
The bottom row panels represent the coupling coefficients $x_{\alpha}$ (by $\lya$ photon scattering, solid line) and $x_{\rm C}$ (by gas collision, dotted line).
The dashed line represents unity. When the coupling coefficients are higher than unity, the spin temperature can decouple from the CMB temperature. The vertical dash-dot line indicates the location of the ionization front at $\chi_{\rm HII} = 0.5$, while the long-dash-dot line indicates the location where $\Tgas = \Tcmb$. In region beyond the long-dash-dot line, if the coupling coefficient is higher than unity, then it would cause absorption, as is the case of the galaxy model.  
}
\label{fig:tb}
\end{center}
\end{figure}

%\begin{figure}
%\begin{center}
%\includegraphics[scale=0.4]{xa.eps}
%\caption{
%Coupling coefficients $x_{\alpha}$ (by $\lya$ photon scattering, solid line) and $x_{\rm C}$ (by gas collision, dotted line) as a function of distance from the central source at redshift $z=10$ for a Pop III star (top panel), a galaxy (middle panel), and a QSO (bottom panel), respectively.  The radius is normalized by Str\"{o}mgren sphere radius $R_{\rm S}$. The dashed line represents unity. When the coupling coefficients is higher than unity, the spin temperature can decouple from the CMB temperature. The vertical dash-dot line indicates the location of the ionization front at $\chi_{\rm HII} = 0.5$, while the long-dash-dot line indicates the location where $\Tgas = \Tcmb$. In region beyond the long-dash-dot line, if the coupling coefficient is higher than unity, then it would cause absorption, as is the case of the galaxy model.  
%}
%\label{fig:xa}
%\end{center}
%\end{figure}

The propagation of the ionization front of a Pop III star, a primordial galaxy, and a QSO are shown in Figure~\ref{fig:ifront}, in comparison with the analytical solution from \cite{Spitzer78}. The position of the ionization front is measured where the ionized hydrogen fraction $\chi_{\rm HII} = 0.5$. The size of the ionized region from the galaxy and QSO follows the analytical one closely, but that of the Pop III star appears to be smaller, owing to helium ionization by Pop III stars. As shown in Figure~\ref{fig:sed}, the SED of Pop III stars peaks around the Lyman limit frequency of HeII, hence a fraction of the ionizing photons is consumed by helium ionization, resulting in a smaller H{\sc ii} region compared to the analytical solution which assumes that all the ionizing photons are absorbed by neutral hydrogen. 

The resulting structures of the ionization, temperatures and differential brightness temperature of the above ionizing sources are shown in Figure~\ref{fig:tb}. The ionization structures of both the Pop III star and QSO show smooth transitions from ionized to neutral state for hydrogen, while the galaxy shows a sharp transition. This is because the high-energy photons of PopIII stars and QSOs have long mean free paths, and they can partially ionize the gas, making the transition smooth. Moreover, the hard SEDs of Pop III stars and QSOs can also produce ionized HeII regions, which is absent in the modeled galaxy without accreting black holes. The size of the ionized He{\sc iii} region is a fraction of $\sim 0.2$ of the H{\sc ii} bubble in Pop III stars and $\sim 0.1$ in QSOs.

\begin{figure*}
\begin{center}
\includegraphics[scale=1.0]{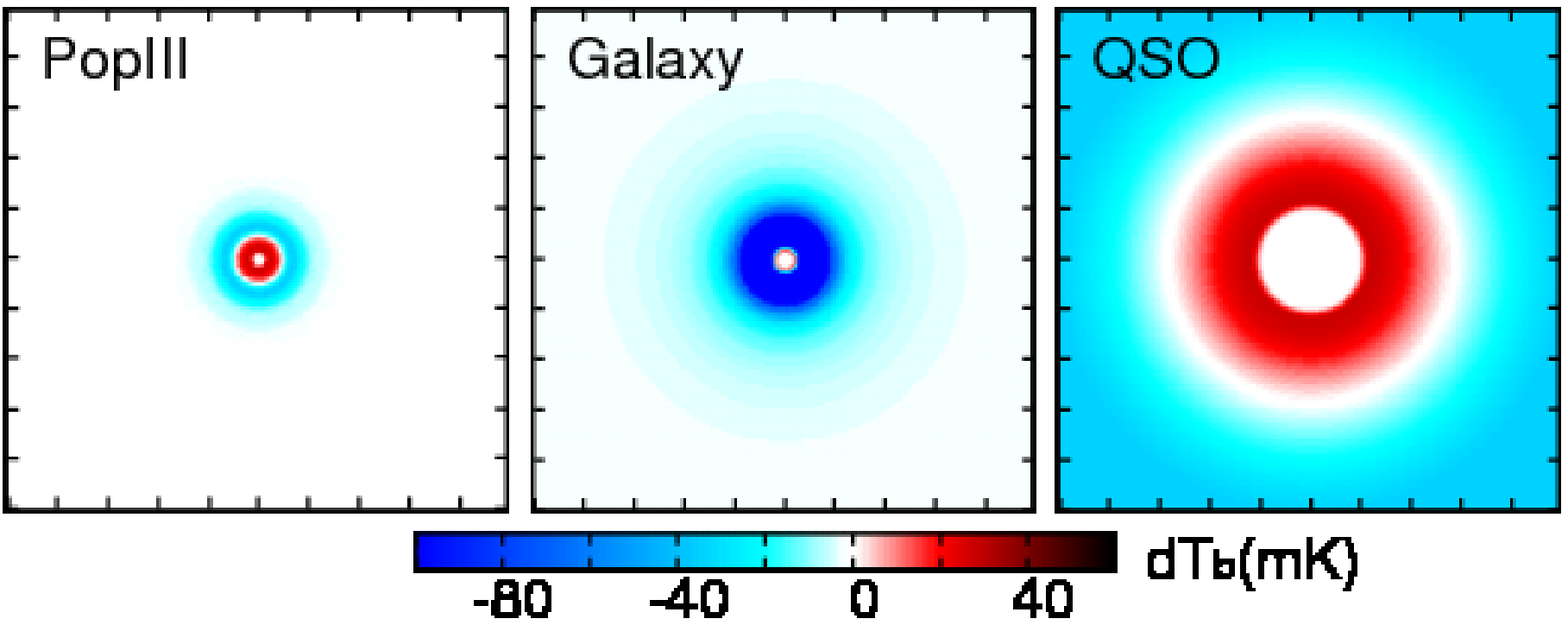}
\caption{
Two-dimensional map of the differential brightness temperature of Pop III stars ($\Mstar = 2\times10^{2}~\rm \Msun$),
galaxy ($\Mstar = 2\times 10^{6}~\rm \Msun$) and QSO ($\Mstar = 2\times10^{8}~\rm \Msun$) at $\tevo=10^{7}~\rm yr$.
 The box size is 500 kpc in physical scale.
}
\label{fig:map}
\end{center}
\end{figure*}

\begin{figure}
\begin{center}
\includegraphics[scale=0.4]{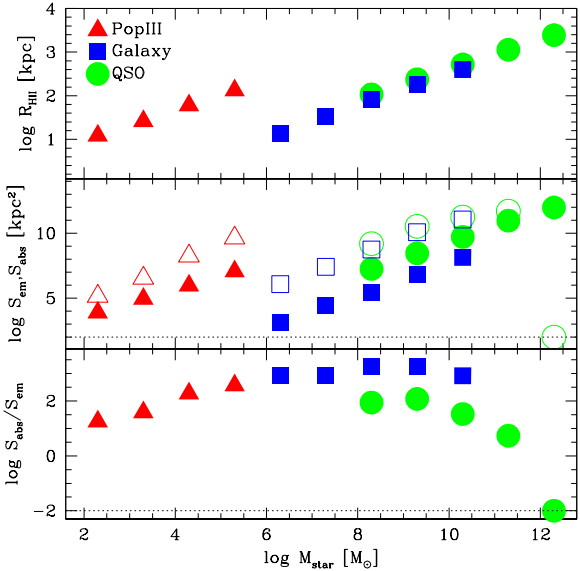}
\caption{Dependence of the 21 cm structures on the source mass for Pop III stars (green triangles), galaxies (red squares), and QSOs (blue circles). 
{\it Upper panel} : size of the ionized bubble $R_{\rm HII}$ at $\tevo=10^{7}~\rm yr$, defined as the position at $\chi_{\rm HII} = 0.5$.
{\it Middle panel} : surface area of emission $S_{\rm em}$ (filled symbols) and absorption ring $S_{\rm abs}$ (open symbols).
{\it Lower panel} : ratio of $S_{\rm abs}$/$S_{\rm em}$ as a function of stellar mass of the source. The dotted lines are artificial lower limits.
}
\label{fig:ring}
\end{center}
\end{figure}

Similarly, the temperatures show different structures from these different sources. The kinetic temperature of the gas, $\Tgas$, shows a smooth transition from a hot ($\sim 10^{4}~\rm K$) to cold ($\sim$ a few K) state around the Pop III star and the QSO, in contrast to the sharp transition in the galaxy, due to partial ionization and heating from photoionization. The spin temperature $\Ts$ shows different patterns in these sources. As indicated in Equation~\ref{eq:ts}, $\Ts$ depends strongly on the coupling efficiency due to $\lya$ scattering and gas collision. If the coupling is strong, $\Ts$ is coupled to $\Tgas$ which is in general different from the CMB temperature $\Tcmb$.   As shown in the second row panels of Figure~\ref{fig:tb}, for the Pop III star, $\Ts$ is weakly coupled to $\Tgas$ within the H{\sc ii} region due to the small number of the $\lya$ scatterings resulted from low number of $\lya$ photons in its SED. As the $\lya$ scattering becomes sparse beyond the H{\sc ii} region, $\Ts$ is decoupled from $\Tgas$ and it takes the value of $\Tcmb$. For the galaxy, $\Ts$ is completely coupled to $\Tgas$ at $R \lesssim 40~\rm kpc$ due to strong $\lya$ scattering, it is thus completely decoupled from $\Tcmb$. For the QSO, $\Ts$ is tightly coupled to $\Tgas$ within $R \lesssim 300~\rm Kpc$ ($\sim$ three times of the size of H{\sc ii} bubble), but outside of which it decouples from $\Tgas$ and becomes $\Tcmb$ due to the sharp decline of $\lya$ scattering. 

As a result of the different ionization and temperature structures around the three sources, the 21 cm signal shows different features, as shown in $\Tb$ in the lower panels in Figure~\ref{fig:tb}. The modeled galaxy shows a narrow emission ring surrounded by an extended absorption trough owing to efficient coupling between $\Ts$ and $\Tgas$. In the case of Pop III star and QSO, since the travel of most $\lya$ photons from the recombination process is confined to the transition region with numerous resonant scatterings within the limited $\tevo=10^{7}~\rm yr$, the $\Tgas$ at transition region is higher than $\Tcmb$, so they both show extended emission in $\Tb$ in the transition region, 
%and the QSO shows only emission in $\Tb$ due to the large size of its H{\sc ii} bubble.
and the QSO shows the extended absorption region due to continuum flux. However, due to the dilution of flux, the absorption signal is weak ($\Tb \lesssim 50$ K). 

The 21 cm structure of QSOs from our model differs from that of \citet{Alvarez10}, which showed extended deep absorption shell in the outer cold region under the assumption of $\Ts = \Tgas$. 
In addition, \citet{Datta12} used the assumption of $\Tb = 4.6\times10^{4} {\rm ~K~cm^{3}} \; n_{\rm HI}({z, \rm comoving})\sqrt{z+1}$, 
which leads to emission even in the region far from QSOs.
 In the case of Pop III stars, the absorption feature from our calculations is shallower than that of \citet{Chen08}. 
This is caused by a smaller $P_{\alpha}$ due to the limited traveling time of $\lya$ photons.
%\textcolor{blue}{which comes from the recombination in H{\sc ii} region and the continuum at between Lyman limit and Ly$\beta$ frequencies. }

To further illustrate the physical process behind the emission and absorption features of the 21 cm signal, we show the radial profiles of the coupling coefficients caused by $\lya$ scatterings, $x_{\alpha}$, and by gas collisions, $x_{\rm C}$, respectively. 
When the coupling coefficient is higher than unity, the spin temperature can decouple from the CMB temperature. As can be seen, $x_{\rm C}$ is around unity in high-temperature region, and becomes very small in region with neutral hydrogen gas. As a result, gas collision has little effect on the change of $\Ts$, the $\Tb$,  in neutral region. On the other hand, $x_{\alpha}$ shows a much higher value than unity even in the outer neutral region, 
and it decreases steeply with distance, because the scattering cross section decreases due to high relative velocity, and $\lya$ photons emitted near sources cannot propagate to outer region within the limited travel time. 
On the other hand, the continuum photons can propagate long distance until they are trapped as Lyman series photons due to the Doppler shift.
Consequently, due to the contribution of $\lya$ photons traveling near the sources, the distribution of $x_{\alpha}$ shows complicated structure, not the simple $r^{-2}$ profile.
In the case of Pop III stars, $x_{\alpha}$ is $\sim 1-100$ in the partially ionized and moderately high  temperature region $\Tgas > \Tcmb$. Therefore, in such a region, $\Ts$ is much higher than $\Tcmb$, resulting in emission in $\Tb$. 
However, at the location of $\Tgas \sim \Tcmb$, $x_{\alpha}$ becomes $\sim 3$, and less than unity in the cold region, leading to the shallower absorption structure.
The $x_{\alpha}$ of the QSO case shows a similar trend with that of the Pop III star. 
%However, since its H{\sc ii} bubble is larger than that of a Pop III star, most of $\lya$ photons cannot reach the point of $\Tgas = \Tcmb$ within the traveling time, therefore, $x_{\alpha}$ falls below unity in high-temperature,  ionized region and sharply drops beyond the Str\"{o}mgren sphere radius $\sim R_{\rm S} (=347~\rm kpc)$, resulting in the no-absorption feature.
However, since the continuum flux is much higher than for Pop III stars, 
these photons propagate a long distance and make the extended absorption region. 
In addition, we can see the $r^{-2}$ profile in the outer region with these continuum photons. 
In the case of the galaxy, $x_{\alpha}$ appears to be very large ($\sim 70$) at the location of $\Tgas = \Tcmb$, and it drops to $\sim 1$ at $R \sim 60~\rm kpc$, resulting in the extended absorption region over  a few hundreds kpc.
%between  $R \sim 20 - 200~\rm kpc$. 

\subsection{The 21 cm Structures}

The resulting 21 cm maps of the Pop III star, galaxy, and QSO  at $z=10$ are shown in Figure~\ref{fig:map}. Clearly, the 21 cm structures of these different sources are distinctively different. The $\Tb$ of the PopIII star shows a ring structure with emission in the inner region and absorption in the outer region. The galaxy shows a very thin emission ring but a deep, extended absorption region, %while the QSO shows extended emission only. 
while the QSO shows an extended emission ring and an outer weak absorption region. 

To further investigate the dependence of 21 cm structures on source properties, we show the emission and absorption structures of the three sources at different stellar masses in Figure~\ref{fig:ring}, as represented by the radius of the ionized bubble $R_{\rm HII}$, the surface area of emission ($S_{\rm em}$) and absorption ($S_{\rm ab}$), and their ratio. 

First, the size of the ionized region increases with the mass of the source, as shown in Figure~\ref{fig:ring} (top panel).  This can be understood since the total number of ionizing photons is simply proportional to the cube-root of the total stellar mass in our models, $R_{\rm HII} \propto M_{\rm star}^{1/3}$.  $R_{\rm HII}$ is in the range of $\sim 24 - 260~\rm kpc$ for PopIII stars, $\sim 27 - 792 ~\rm kpc$ for galaxies, and $\sim 0.2 - 4.9 ~\rm Mpc$ for QSOs. The size of the region ionized by QSOs is consistent with that from \cite{Wyithe05}. Pop III stars have a higher ionizing power because their  hard SEDs produce more ionizing photons and higher effective temperature $\sim 10^{5}~\rm K$. 
%Since the $\lya$ photons cannot be scattered in the ionized region, 
Of course, the ionized region would appear as a ``zero-signal hole" ($\Tb \sim 0~\rm mK$) in the 21 cm structure as shown in Figure~\ref{fig:map}. 

The middle panel of Figure~\ref{fig:ring} shows the surface areas of the emission ($S_{\rm em}$) and absorption ($S_{\rm abs}$) regions in a two dimensional slice containing the central sources. For all models, $S_{\rm em}$ monotonically increases with the stellar mass of the source; the $S_{\rm abs}$, however,  is clearly different depending on the source type and the size of ionized region. The $S_{\rm abs}$ of galaxies are much higher than the $S_{\rm em}$, which produces the strong absorption trough as seen in Figure~\ref{fig:map}. 
In the case of QSOs, $S_{\rm abs}$ decreases with increasing stellar mass at $M_{\rm star} \gtrsim 10^{10}~\Msun$, 
since even continuum photons cannot propagate to the outside cold, neutral gas region in the limited time. 
The ratio between $S_{\rm abs}$ and $S_{\rm em}$ is shown in the lower panel of Figure~\ref{fig:ring}. The ratio is in the range of $\sim 20 - 370$ for Pop III stars, $\sim 830 - 1850$ for the galaxies, and  $\sim 0 - 120$ for the QSOs.

These results demonstrate that Pop III stars, the first galaxies, and the first QSOs have clearly different 21 cm structures, owing to different temperature and ionization structures resulting from different photon SEDs and propagation of $\lya$ photons in these regions. 

\begin{figure}
\begin{center}
\includegraphics[scale=0.4]{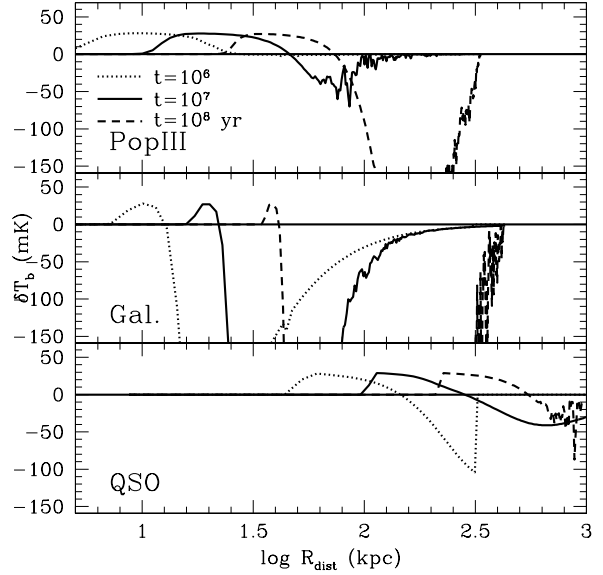}
\caption{
The evolution of the differential brightness temperature of a Pop III star, a galaxy, and a QSO at time $\tevo=10^{6}, 10^{7}$ and $10^{8}~\rm yr$, respectively. The mass of these three sources is $\Mstar = 2\times10^{2}, 2\times10^6, 2\times10^8\, ~\rm \Msun$, respectively. 
}
\label{fig:tbtime}
\end{center}
\end{figure}

\begin{figure}
\begin{center}
\includegraphics[scale=0.4]{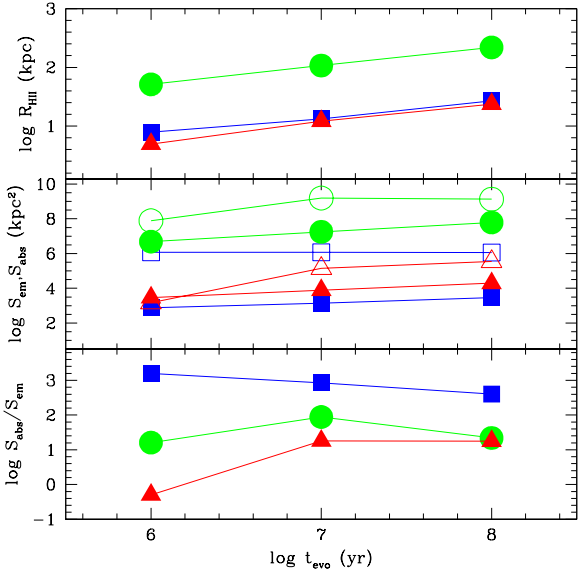}
\caption{
The evolution of the size of the ionizing bubble $R_{\rm HII}$, surface area of emission $S_{\rm em}$ and absorption $S_{\rm abs}$, and their ratio $S_{\rm abs}$/$S_{\rm em}$, at different times for Pop III stars (red triangles), galaxies (blue squares), and QSOs (green circles). The masses of these three different sources are $\Mstar = 2\times10^{2}, 2\times10^6, 2\times10^8\, ~\rm \Msun$, respectively. The symbols are the same as in Figure~\ref{fig:ring}. 
}
\label{fig:ring_time}
\end{center}
\end{figure}

\begin{figure}
\begin{center}
\includegraphics[scale=0.4]{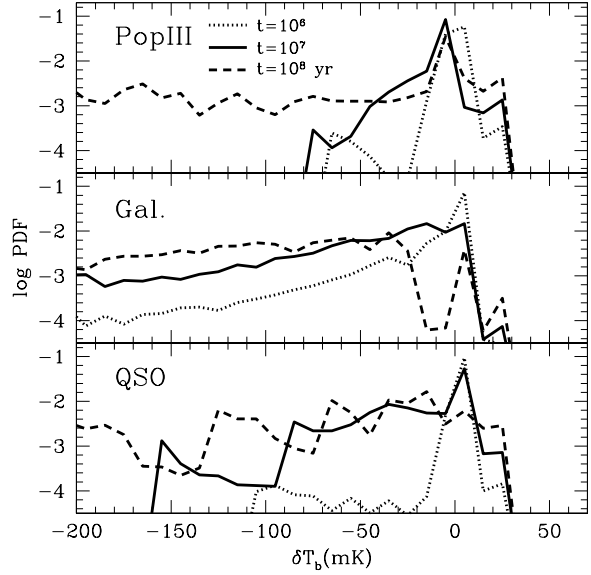}
\caption{
Probability distribution function (PDF) of the 21 cm emission volume normalized by the one of $5\times$ Str\"{o}mgren sphere. 
We derived the normalized emission volume of each stellar mass, and stacked the PDFs.
}
\label{fig:pdf}
\end{center}
\end{figure}

\subsection{Time Evolution of the 21 cm Structure}

The evolution of the $\Tb$ signal at different times is shown in Figure~\ref{fig:tbtime}. As the ionizing bubble grows with time, the $\lya$ photons propagate into the more extended cold region through scattering, resulting in the stronger absorption signal. At $\tevo = 10^{6}~\rm yr$, the $\Tb$ of Pop III stars %and QSO 
has no absorption region. At $\tevo = 10^{8} ~\rm yr$, all simulations show 21 cm absorptions.  Note that for Pop III stars, the evolution of ionization may not last $10^{8}~\rm yr$ due to their short lifetimes and fast metal enrichment \citep[e.g.,][]{Wise12a}. However, since the recombination time scale of IGM is longer than $10^{8}~\rm yr$, the $\lya$ photons can travel in the relic H{\sc ii} region and the cold region outside, which may produce similar absorption features as in Figure~\ref{fig:tbtime}. 
 
Figure~\ref{fig:ring_time} shows the evolution of the size of the ionizing bubble $R_{\rm HII}$, surface area of emission $S_{\rm em}$ and absorption $S_{\rm abs}$, and their ratio $S_{\rm abs}/S_{\rm em}$, at different time $\tevo=10^{6}, 10^{7}$ and $10^{8}~\rm yr$  for Pop III stars, galaxies, and QSOs.   
$R_{\rm HII}$ increases with $\tevo$, $R_{\rm HII} \propto (1-{\rm exp}(-\tevo/t_{\rm rec}))^{1/3}$, where $t_{\rm rec}$ is recombination time scale. 
The $S_{\rm em}$ and $S_{\rm abs}$ of galaxies and QSOs show the slow increase with time, and the ratio of the two becomes nearly constant after $\sim 10^{7}~\rm years$. 
In the case of Pop III stars, however, $S_{\rm abs}$ increases dramatically after $\sim 10^{7}~\rm years$, leading to a strong absorption feature.
This is because $\lya$ photons from the recombination process are the dominant source of $\lya$ photon scatterings,
and they are trapped near the source at $t \lesssim 10^{7}$ yr and can propagate to the outer cold region at $t \sim 10^{8}$ yr.

The probability distribution function (PDF) of the 21 cm emission volume is shown in Figure~\ref{fig:pdf}.
As time evolves, more $\lya$ photons propagate to cold gas region, producing the extended tail of the PDF with the absorption signal.
At $\tevo = 10^{7}~\rm yr$, the PDF of Pop III stars is confined to $\Tb \gtrsim -80$ mK, and it extends to $\Tb < -100$ mK at $10^{8}~\rm yr$. 
In the case of the QSO, the PDF has the more extended tail at $\Tb \lesssim -160$ mK at $\tevo = 10^{7}~\rm yr$ due to the continuum photons.

%%%%%%%%%%Figure

\subsection{Detectability}

\begin{figure}
\begin{center}
\includegraphics[scale=0.4]{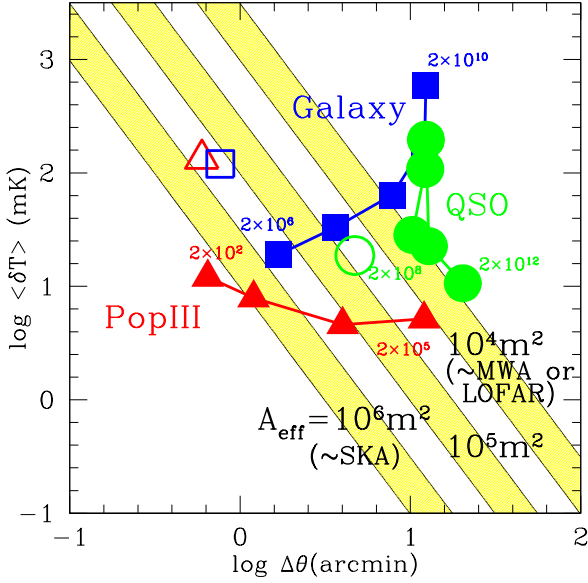}
\caption{
Detectability of the 21 cm signal from Pop III stars, the first galaxies, and the first quasars with upcoming missions MWA, LOFAR, and SKA. 
The filled symbols represent the %mean 
$\Tb$ at $z=10$ with different stellar masses as shown in Figure~\ref{fig:ring}, which is estimated in the spherical top hat beam with size of the distance from the source to the 
outer edge of $|\Tb| = 1~\rm mK$. 
%2D $\Tb$ maps with widths of $\Delta \nu = 1, 2, 3$ and $4$ MHz are used to estimate the maximum and minimum values of the elliptical shades. 
For each stellar mass,  we calculate 2D $\Tb$ maps with widths of $\Delta \nu = 1, 2, 3$ and $4$ MHz, and take the mean values of the four slices. 
With increasing stellar mass, the symbols shift to higher $\Delta \theta$ due to the larger sizes of the H{\sc ii} region and the 21 cm ring. 
The open symbols are the mean $\Tb$ of the lowest stellar masses for each source at $z=20$.
The yellow shaded regions indicate the sensitivity for $z=10$ by upcoming facilities with $A_{\rm eff} = 10^{4}, 10^{5}$ and $10^{6}~\rm m^{2}$
estimated by equation (\ref{eq:sensitivity}), while its width corresponds to the integration time from $100$ to $1000$ hours 
with $\Delta \nu = 1~\rm MHz$. 
}
\label{fig:detectability}
\end{center}
\end{figure}

The detectability of the 21 cm emission from the first luminous objects with future instruments is of great interest. 
Here we calculate the detectability of the 21 cm emission from Pop III stars, first galaxies and quasars with upcoming observatories such as MWA, LOFAR, and SKA. The noise per resolution element  is estimated by \citet{Furlanetto09} as follows,
\begin{eqnarray}
\label{eq:sensitivity}
\Delta T &\sim 20~{\rm mK}~\left(\frac{10^{4}~\rm m^{2}}{A_{\rm eff}} \right) \left( \frac{10'}{\Delta \theta} \right) \cr
& \times \left( \frac{1+z}{10}\right)^{4.6} \left( \frac{\rm MHz}{\Delta \nu} \frac{100~\rm hr}{t_{\rm int}} \right)^{1/2},
\end{eqnarray}
where $A_{\rm eff}$ is the effective collect area, $\Delta \theta$ is the angular diameter, $\Delta \nu$ is the band width, 
and $t_{\rm int}$ is the integration time. 

For the redshifted 21-cm line at $z\sim 10$, the contribution from the polarized Galactic synchrotron foreground is dominant over the system temperature,
and it has $T \sim 180 (\nu / 180\; \rm MHz)^{-2.6}$. The sensitivity curves are shown in Figure~\ref{fig:detectability}. 
The yellow shade with $A_{\rm eff} = 10^{4}~\rm m^{2}$ roughly corresponds to the sensitivity of MWA and LOFAR,
while that with $A_{\rm eff} = 10^{6}~\rm m^{2}$ corresponds to the sensitivity of SKA.

Here, we measure the size of the emission or absorption spheres accompanying the modeled sources with boundaries of  $|\Delta T_{\rm b}| = 1~\rm mK$.
We calculate the $\delta T_{\rm b}$ map with thickness of $\Delta \nu = 1, 2, 3$ and $4~\rm MHz$ around central sources,
and we estimate $<\delta T>$  by taking the absolute value of the spatial mean of $\delta T_{\rm b}$ at $t_{\rm evo} = 10^{7}~\rm yr$.
The $<\delta T>$ is not sensitive to the different thicknesses, and we take the mean values of the $<\delta T>$ for the four different thicknesses.  
The results of this are shown in Figure~\ref{fig:detectability}. Different symbols represent the results of the different stellar masses as shown in Figure~\ref{fig:ring}.
The size $\Delta \theta$ increases with the stellar mass and the size of H{\sc ii} bubbles. 
At $z=10$, Pop III stars show $<\delta T> \sim 5 - 12~\rm mK$, while galaxies have $<\delta T> \sim 19 - 584~\rm mK$ and QSOs have $<\delta T> \sim 28 - 196~\rm mK$. 
In Pop III case, there are absorption spheres comparable to emission ones. Hence the net signal becomes small due to the offset. 
For the QSO case, the emission spheres are relatively small to the central zero-signal holes, hence the net signal decreases. 
In addition, the net signal decreases for massive QSOs from the offset between emission and absorption, 
because the absorption region becomes small due to the limited propagation distance of photons. 
On the other hand, in the case of galaxies, there are deep extended absorption spheres relative to the central holes and inner emission spheres.
Thus, galaxies appear to have much stronger 21 cm signals than the Pop III stars and QSOs. 
The H{\sc ii} regions around luminous QSOs and massive galaxies can be detected by MWA and LOFAR with $\sim 100$ hours integration. 
%For faint QSOs $\Mstar \sim 10^{10}~\Msun$, it may need $\sim 1000$ hours integration by MWA and LOFAR.
For low-mass galaxies, SKA or an $A_{\rm eff}=10^{5}~\rm m^{2}$ class telescope will be needed to detect the H{\sc ii} regions around them.
In the case of Pop III stars with typical mass $M_{\rm star} \lesssim 1000~\Msun$, however, it appears to be difficult to detect the H{\sc ii} region even with SKA. 

At $z=20$,  the $<\delta T>$ of a Pop III star with $2 \times 10^{2}~\Msun$ and a galaxy with $2 \times 10^{6}~\Msun$ is enhanced by factor $\sim 6-27$
due to the decrease of the sizes of their H{\sc ii} regions and the higher gas density.
On the hand, the sensitivity gets lower from $z=10$ to $20$ by factor $\sim 20$. Hence, the detectability of Pop III stars and low-mass galaxies does not change significantly. 
In the case of the QSO with $M_{\rm star}=2\times10^{8}~\Msun$, the continuum radiation is significantly diluted in cold neutral region due to the larger H{\sc ii} region. 
The increase of $S_{\rm ab}/S_{\rm em}$ is not large compared to Pop III stars, and hence $<\delta T>$ does not change as much.
The redshift evolution of the 21 cm structure is also discussed in Section~\ref{sec:discussion}

Our results suggest that  LOFAR, MWA or SKA may be able to detect the 21 cm signal around galaxies or QSOs, and may be able to distinguish the natures of sources through the difference of emission or absorption and the size. Here, we optimistically assume that the beam size of observations is comparable to the size of the outer emission/absorption shells. In practice, the beam size can be larger than the one used in our models. 
As a result, the net signal may decrease with the beam size. However, the decrease will follow the same gradient of the sensitivity curve $\Delta \theta^{-2}$, hence it may not significantly affect the detectability in the above discussion. 
In addition, note that we have ignored UV background radiation in this work. 
If there is UV background radiation and $x_{\alpha}$ becomes $\ge 1$ at everywhere, 
even IGM far from sources will show emission or absorption \citep[e.g.,][]{Baek09}.
This leads to a uniform 21 cm signal over the whole sky, and makes the identification of individual sources difficult. 
For such a case, only high-redshift QSOs can be identified via detection of giant holes in 21 cm sky map \citep{Vonlanthen11}.
In addition, the X-ray background made by black holes and supernovae affects the 21 cm signal by heating the IGM \citep{Furlanetto06, Yajima14}.
The strength of UV and X-ray background depends on the cosmic star formation history and the formation of massive black holes, which are still under the debate.
Recent observation of high-redshift galaxies have suggested that the cosmic star formation rate steeply decreases with increasing redshift at $z \gtrsim 8$ \citep[e.g.,][]{Oesch13}.
If so, $x_{\alpha}$ is unlikely to be much higher than unity at $z\sim10$.   
Therefore,  the 21 cm structure around sources at $z \lesssim 20$ should be studied along with formation of galaxies and black holes in a large volume. 
We will investigate the 21 cm structure's relationship with galaxy formation throuth cosmological simulation in future work. 

%----------------------------------------------------------------------
%
% Section 4:  Discussion
%
%----------------------------------------------------------------------

\section{Discussion}
\label{sec:discussion}

\begin{figure}
\begin{center}
\includegraphics[scale=0.4]{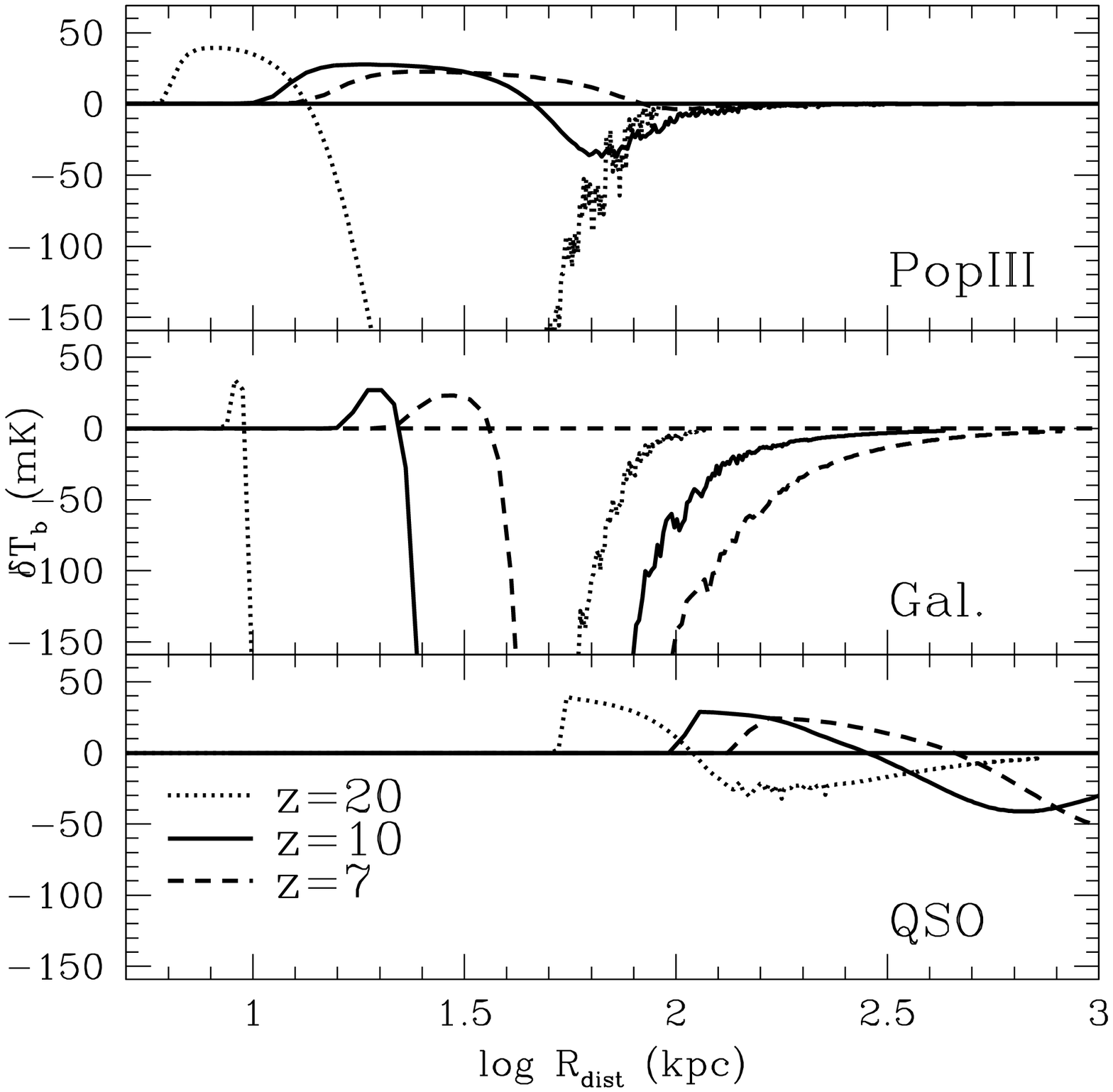}
\caption{
The differential brightness temperature of Pop III stars ($\Mstar = 10^{2}~\rm \Msun$),
galaxy ($\Mstar = 10^{6}~\rm \Msun$) and QSO ($\Mstar = 10^{8}~\rm \Msun$)
at $z=7, 10$ and $20$. The snap shots at $\tevo=10^{7}~\rm yr$.
}
\label{fig:tbz}
\end{center}
\end{figure}

In this work, we focus on the 21 cm signal at $z=10$. 
However, the size of the ionizing bubble and the propagation distance of $\lya$ change at different redshifts. 
We test the change of $\Tb$ structure at $z=7$ and $20$ with same source properties.
Figure~\ref{fig:tbz} shows the $\Tb$ structure with $t_{\rm evo} = 10^7 ~\rm yr$ at different redshifts.
The size of ionized bubbles decreases with increasing redshift, 
while the ionizing front becomes closer to the Str\'{o}mgren radius in the same propagation time.
As a result, the expansion rate of the ionizing front is estimated by 
\begin{equation}
\frac{R^{'}_{\rm I}}{R_{\rm I}} = \left( \frac{1+z'}{1+z} \right)^{-2} 
\left[ \frac{1-e^{- A (1+z')^{3}}}{1-e^{- A (1+z)^{3}}} \right]^{1/3}
\end{equation}
where $A=\tevo \alpha_{\rm B} n^{-1}_{0}$.
Hence, the size ratio of $z=7 \;(20)$ to $z=10$ is $\sim 1.2 \;(0.5)$.
For Pop III stars at $z=20$, there is a deep extended absorption area.
With decreasing redshift, the absorption signal becomes shallower,
because the distance to the cold-neutral region becomes large, leading to a larger travel time to there.
In the case of galaxies, the 21 cm structure is almost the same, however, the absorption area is relatively small at lower redshift, 
because the H{\sc ii} bubble is bigger and the transition from ionized to neutral becomes more smooth due to the lower density.
Note that the QSO at $z=20$ shows a clear absorption signal, despite the fact that the H{\sc ii} region is much larger than that of Pop III star at $z=7$ which does not show a clear absorption signal.
This is due to the higher $\lya$ luminosity of the continuum spectrum than the Pop III star.

%----------------------------------------------------------------------
%
% Section 5:  Summary
%
%----------------------------------------------------------------------

\section{SUMMARY}

In this paper, we investigate the redshifted 21 cm signal from different ionizing sources by combining idealized models of the first stars, the first galaxies, and quasars with radiative transfer simulations that include both ionization of neutral hydrogen and resonant scattering of $\lya$ photons.  We find that the 21 cm signal depends significantly on the source SED and stellar mass. The Pop III stars and quasars produce a smooth transition from an ionized and hot state to a neutral and cold one, owing to their hard SED with abundant ionizing photons, in contrast to the sharp transition in galaxies. The H{\sc ii} bubble size is typically $\sim 20 ~\rm kpc$ for PopIII stars, $\sim 100~\rm kpc$ for galaxies, and $\sim 5~\rm Mpc$ for QSOs. Furthermore,  $\lya$ scattering plays a dominant role in producing the 21 cm signal as it determines the relation between hydrogen spin temperature and gas kinetic temperature. The $\lya$ photons can produce both emission and absorption region around the small H{\sc ii} bubbles of PopIII stars, extended absorption region around the first galaxies, and extended emission around QSOs. 

We predict that future surveys from large radio arrays such as MWA, LOFAR and SKA may be able to detect the 21 cm signals of primordial galaxies and quasars, but not likely Pop III stars due to their small angular diameter.

These results are based on idealized models of the first stars, the first galaxies, and quasars. We will investigate the 21 cm structures in more realistic scenarios including inhomogeneous density, metal enrichment and black hole growth by using cosmological simulations in future works.

%----------------------------------------------------------------------
%
% Acknowledge
%
%----------------------------------------------------------------------
\section*{Acknowledgments}
Support from NSF grants AST-0965694, AST-1009867 (to YL), and AST-0807075 (to TA) is gratefully acknowledged.  We acknowledge the Research Computing and Cyberinfrastructure unit of Information Technology Services at The Pennsylvania State University for providing computational resources and services that have contributed to the research results reported in this paper (URL: http://rcc.its.psu.edu). 
The Institute for Gravitation and the Cosmos is supported by the Eberly College of Science and the Office of the Senior Vice President for Research at the Pennsylvania State University.

%----------------------------------------------------------------------
%
% References
%
%----------------------------------------------------------------------
%\begin{thebibliography}{99}

%\bibliographystyle{mn}

%\bibliography{mn-jour,HY}

%\end{thebibliography}

\label{lastpage}

\end{document}